\documentclass[11pt,a4paper]{article}
\usepackage{times}
\usepackage[T1]{fontenc}
\usepackage[latin1]{inputenc}
\usepackage[francais]{babel}
\usepackage{fancyhdr}
\usepackage{amssymb}
\usepackage{undertilde}
\usepackage{amsmath}
\usepackage{setspace}
\usepackage{graphicx}
\usepackage{wrapfig}
\usepackage{titlesec}
\titleformat{\section}
  {\normalfont\fontsize{16}{20}\bfseries}{\thesection}{1em}{}
\titleformat{\subsection}
  {\normalfont\fontsize{16}{20}\bfseries}{\thesubsection}{1em}{}
\titlespacing*{\section}
{0pt}{2.ex plus 1ex minus .2ex}{.0ex}
\titlespacing*{\subsection}
{0pt}{0.ex}{.0ex}

\begin{document}

\begin{center}
\begin{spacing}{2.05}
{\fontsize{20}{20}
\bf
Modélisation de la flexion d'une lame de polymère électroactif
}
\end{spacing}
\end{center}
\vspace{-1.25cm}
\begin{center}
{\fontsize{14}{20}
\bf
M. TIXIER\textsuperscript{a}, J. POUGET\textsuperscript{b}\\
\bigskip
%\vspace{0.75cm}
}
{\fontsize{12}{20}
a. Département de Physique, Université de Versailles Saint Quentin,
45, avenue des Etats-Unis, F-78035 Versailles, France;
mireille.tixier@uvsq.fr\\
b. Sorbonne Université, UPMC Univ. Paris 6, UMR 7190, Institut Jean le Rond d'Alembert, F-75005 Paris, France; CNRS, UMR 7190, Institut Jean le Rond d'Alembert, F-75005 Paris, France; pouget@lmm.jussieu.fr\\
} 
\end{center}

\vspace{10pt}

{\fontsize{16}{20}
\bf
R\'esum\'e :
}
\medskip

\textit{Un polymère électro-actif ionique (le Nafion par exemple) peut être utilisé comme capteur ou comme actionneur.
Pour ce faire, on place une fine couche de ce matériau saturé d'eau entre deux électrodes. La saturation en eau 
entraîne une dissociation quasi complète du polymère et la libération de cations de petite taille. L'application 
d'un champ électrique perpendiculaire à la lame provoque la flexion de celle-ci. Inversement, le fléchissement de 
la lame fait apparaître une différence de potentiel entre les électrodes. Ce phénomène fait intervenir des 
couplages multiphysiques de type électro-mécano-chimiques. Nous avions précédemment modélisé ce système et déterminé 
ses lois de comportement grâce à la thermodynamique des processus irréversibles linéaires. \newline}
\textit{\noindent Nous avons appliqué ce modèle au cas d'une lame de PEA encastrée - libre soumise à une différence 
de potentiel continue entre ses deux faces (cas statique). Les efforts appliqués et la flèche sont calculés en
utilisant un modèle de poutre en grands déplacements. Nous avons également étudié la force qu'il faut exercer 
sur l'extrémité libre pour l'empêcher de se déplacer (force de blocage). \newline}
\textit{\noindent Les simulations numériques ont été effectuées dans le cas du Nafion. Nous avons tracé les profils 
de concentration en cations, de pression, de potentiel et de champ électriques dans l'épaisseur de la lame. 
Ces grandeurs, quasiment constantes dans la partie centrale de la lame, varient de façon drastique au voisinage 
des électrodes. Les valeurs obtenues pour la flèche et de la force de blocage sont en bon accord avec les données 
expérimentales publiées dans la littérature. Nous retrouvons une variation linéaire de ces deux grandeurs avec le 
potentiel électrique imposé; la flèche varie comme le carré de la longueur, et la force de blocage est 
proportionnelle à la largeur et à l'inverse de la longueur.}

\vspace{20pt}

{\fontsize{16}{20}
\bf
Abstract :
}
\bigskip

\textit{Ionic electro-active polymer (Nafion for example) can be used as sensor or actuator. To this end, 
a thin film of the water-saturated material is sandwiched between two electrodes. Water saturation causes a 
quasi-complete dissociation of the polymer and the release of small cations. The application of an electric 
field across the thickness results in the bending of the strip. Conversely, a voltage can be detected between the two
electrodes when the strip is bent. This phenomenon involves multiphysics couplings of electro-mechanical-chemical 
type. We have previously modeled this system and determined its constitutive equations using the thermodynamics 
of linear irreversible processes. \newline}
\textit{\noindent We applied this model to the case of a cantilevered PEA strip subjected to a continuous voltage 
between its two faces (static case). The applied forces and the tip displacement are calculated using a beam model 
in large displacements. We have also studied the force to be exercised on the free end to prevent its displacement 
(blocking force). \newline}
\textit{\noindent Numerical simulations were performed in the case of Nafion. We have drawn the profiles of
cations concentration, pressure, electric field and potential in the thickness of the strip. These quantities, 
which are almost constant in the central part of the strip, vary drastically near the electrodes. The  obtained 
values of the tip displacement and the blocking force are in good agreement with the experimental data published 
in the literature. These two quantities are linear functions of the imposed electrical potential; the tip 
displacement varies as the length square, and the blocking force is proportional to the width and inversely 
proportional to the length.}

\vspace{28pt}

{\fontsize{14}{20}
\bf
Mots clefs : Electro-active polymers - Multiphysics coupling - Polymer mechanics - Nafion - Smart materials
}
%\bigskip

\section{Introduction}
\medskip
Les polymères électroactifs sont des matériaux innovants pouvant avoir de multiples applications : ils peuvent servir
d'actionneurs pour la confection de micromanipulateurs, de micro-robots ou de micro-pompes. On envisage aussi de les 
utiliser dans le domaine médical comme muscles artificiels, dans le domaine de la bio-inspiration (conception 
d'ailes battantes pour les micro-drones inspirées du vol des insectes), ou encore comme capteur
ou pour la récupération d'énergie.

Nous nous sommes plus particulièrement intéressés à un composite ionique (IPMC) constitué d'une lame de polymère 
ionique (Nafion ou Flemion) recouverte sur ses deux faces d'une fine couche de métal (or ou platine) servant 
d'électrodes. 
Le Nafion est un polyélectrolyte composé d'un squelette hydrophobe de polytétrafluoroéthylène sur lequel sont greffées 
des chaînes pendantes terminées par des groupes sulfoniques hydrophiles. La lame de polymère est saturée d'eau, 
ce qui provoque une dissociation complète des groupes sulfoniques et la libération dans l'eau de cations de petite
taille ($H^{+}$, $Li^{+}$ ou $Na^{+}$), les anions restant fixés sur le squelette. Lorsque l'on applique un champ
électrique perpendiculaire à la 
lame, les cations migrent vers l'électrode négative (cathode), entraînant avec eux l'eau par un phénomène d'osmose. Il 
en résulte un gonflement du polymère du côté de la cathode et une contraction sur l'autre face, ce qui provoque le 
fléchissement de la lame vers l'anode. Une lame de $200\;\mu m$ d'épaisseur de quelques centimètres de long fléchit
ainsi de quelques millimètres en une seconde sous l'action d'une différence de potentiel de quelques volts
\cite{nemat2000}. Le phénomène met donc en jeu des couplages multiphysiques.

Nous avions précédemment modélisé ce système par un milieu poreux déformable (les chaînes polymères chargées
négativement) dans lequel s'écoule une solution ionique formée par l'eau et les cations. L'ensemble est 
assimilé à un milieu continu. 
La thermodynamique des processus irréversibles nous a permis d'en déduire les lois de comportement régissant ce
système. Les principales hypothèses et équations de ce modèle sont rappelées dans le second paragraphe.
Nous avons appliqué ce modèle à la flexion d'une lame encastrée à l'une de ses extrémités dans les deux cas suivants : 
l'autre extrémité est soit libre, soit maintenue fixe par un effort tranchant (force de flocage). Le système 
présentant des déformations de grande amplitude, notamment dans le premier cas, nous avons utilisé un modèle 
de poutre en grands déplacements (paragraphe 3).

Les résultats des simulations sont présentés dans le paragraphe 4. Nous avons tracé les profils du potentiel et du 
champ électriques, de la concentration en cations et de la pression dans l'épaisseur de la lame. Nous avons également 
étudié les variations de la flèche et de la force de blocage avec les caractéristiques géométriques de la lame et avec 
le potentiel électrique imposé.

Nos conclusions sont détaillées dans le paragraphe 5.

\section{Modélisation du polymère}

\subsection{Hypothèses}
\medskip
Notre modélisation est basée sur la themomécanique des milieux continus. Les chaînes polymères chargées négativement 
sont assimilées à un milieu poreux déformable, homogène et isotrope. Le matériau est représenté par la superposition
de trois systèmes mobiles les uns par rapport aux autres : le solide poreux, le solvant (l'eau) et les cations. 
Les phases solide et liquide (eau + cations) sont séparées par une interface. On néglige la gravité et l'induction
magnétique. Les différentes phases sont supposées incompressibles et la solution diluée. On admet en outre que les
déformations du solide sont petites. 

Notre modèle est basé sur un modèle à gros grains développé pour les mélanges à deux constituants \cite{Ishii06}. Les 
équations de conservation sont tout d'abord écrites à l'échelle microscopique pour chaque phase et pour les 
interfaces. 
A l'échelle macroscopique, on définit un volume élémentaire représentatif contenant les deux phases. Les équations 
macroscopiques du matériau sont déduites des équations microscopiques par un processus de moyenne utilisant des 
fonctions de présence pour chaque phase.

\subsection{Equations de bilan}
\medskip
On obtient ainsi les équations de bilan de la masse (\ref{m}), de la quantité de mouvement (\ref{CQ}), de l'énergie 
(\ref{E}), de la charge électrique (\ref{CC}) et les équations de Maxwell (\ref{maxwell}) 
relatives au matériau complet \cite{Tixier1} :
\begin{equation}
\frac{\partial \rho }{\partial t}+div\left( \rho \overrightarrow{V}\right)=0 \label{m}
\end{equation}%
\begin{equation}
\rho \frac{D\overrightarrow{V}}{Dt}=\overrightarrow{div}\utilde{\sigma }%
+\rho Z\overrightarrow{E}  \label{CQ}
\end{equation}%
\begin{equation}
\rho \frac{D}{Dt}\left( \frac{E}{\rho }\right) =div\left( \sum\limits_{phases}%
\utilde{\sigma}_{k}\cdot \overrightarrow{V_{k}}\right) -div\overrightarrow{Q} \label{E}
\end{equation}%
\begin{equation}
div\overrightarrow{I}+\frac{\partial \rho Z}{\partial t}=0 \label{CC}
\end{equation}%
\begin{equation}
\begin{tabular}{lll}
$\overrightarrow{rot}\overrightarrow{E}=\overrightarrow{0}\qquad \qquad $ & $%
div\overrightarrow{D}=\rho Z\qquad \qquad $ & $\overrightarrow{D}%
=\varepsilon \overrightarrow{E}$%
\end{tabular} \label{maxwell}
\end{equation}%
où $\rho$ est sa masse volumique, $\overrightarrow{V}$ sa vitesse, $\utilde{\sigma }$ son tenseur des contraintes, $Z$ 
sa charge électrique massique, $E$ son énergie volumique totale, $\varepsilon$ sa permittivité diélectrique, 
$\overrightarrow{E}$ le champ électrique, $\overrightarrow{Q}$ le flux de chaleur conductif, $\overrightarrow{I}$ 
la densité volumique de courant et $\overrightarrow{D}$ l'induction électrique. Dans ces relations, $\frac{D}{Dt}$ 
désigne la dérivée matérielle, c'est à dire une dérivée en suivant le mouvement des différents constituants.

La relation de Gibbs s'écrit \cite{Tixier2} :
\begin{equation}
T\frac{d}{dt}\left( \frac{S}{\rho }\right) =\frac{d}{dt}\left( \frac{U}{\rho 
}\right) +p\frac{d}{dt}\left( \frac{1}{\rho }\right)
-\sum\limits_{constituants}\mu _{k}\frac{d}{dt}\left( \frac{\rho _{k}}{\rho }%
\right) -\frac{1}{\rho }\utilde{\sigma ^{e}}^{s}:\utilde{grad}%
\overrightarrow{V}  \label{Gibbs}
\end{equation} 
où $T$ est la température absolue, $S$ et $U$ l'entropie et l'énergie interne volumiques, $p$ la pression, $\mu _{k}$ 
les potentiels chimiques des constituants et $\utilde{\sigma ^{e}}^{s}$ la partie symétrique du tenseur des 
contraintes d'équilibre. $\frac{d}{dt}$ désigne ici la dérivée particulaire barycentrique, c'est à dire en suivant le 
mouvement du barycentre des constituants.

\subsection{Lois de comportement}
\medskip
En combinant les lois de conservation et la relation de Gibbs, on peut déterminer la fonction de dissipation 
du matériau \cite{Tixier2}. La thermodynamique des processus irréversibles linéaires permet alors d'identifier 
les flux et les forces généralisées associées et d'en déduire les lois de comportement du matériaux. On obtient une 
loi de Fourier généralisée, une loi de Darcy généralisée (\ref{Darcy}) et une loi de Nernst-Planck (\ref{Nernst}) :
\begin{equation}
\overrightarrow{V_{4}}-\overrightarrow{V_{3}}\simeq -\frac{K}{\eta\phi
_{4}}\left[ \overrightarrow{grad}p-\left(C F - \rho _{2}^{0} Z_{3}\right) 
\overrightarrow{E}\right] \label{Darcy}
\end{equation}%
\begin{equation}
\overrightarrow{V_{1}}=-\frac{D}{C}\left[ \overrightarrow{grad}C-\frac{%
Z_{1}M_{1}C}{RT}\overrightarrow{E}+\frac{Cv_{1}}{RT}\left( 1-\frac{M_{1}}{%
M_{2}}\frac{v_{2}}{v_{1}}\right) \overrightarrow{grad}p\right] +%
\overrightarrow{V_{2}} \label{Nernst}
\end{equation}%
où $\eta$ désigne la viscosité dynamique du solvant, $\phi_{4}$ la fraction volumique de la solution, $K$ la 
perméabilité intrinsèque du solide, $D$ le coefficient de diffusion de masse des cations, $C$ la concentration molaire 
en cations, $F=96487~C~mol^{-1}$ la constante de Faraday, $\rho _{2}^{0}$ la masse volumique du solvant, 
$M_{k}$ la masse molaire, $v_{k}$ le volume molaire partiel et $R=8,31 J~K^{-1}$ la constante universelle des gaz 
parfaits. Les indices $1$, $2$, $3$ et $4$ sont relatifs respectivement aux cations, au solvant, au solide et à 
la solution (solvant + cations).
En supposant qu'en statique, le matériau vérifie la loi de Hooke, la loi rhéologique s'écrit :
\begin{equation}
\utilde{\sigma }=\lambda \left( tr\utilde{\epsilon }\right) \utilde{1}+2G%
\utilde{\epsilon }+\lambda_{v}\left( tr \dot{\utilde{\epsilon }} \right)
 \utilde{1}+2\mu_{v}\dot{\utilde{\epsilon }} \label{rheo}
\end{equation}%
où $\utilde{\epsilon }$ désigne le tenseur des déformations, $\lambda$ le premier coefficient de Lamé, $G$ le module 
de cisaillement et $\lambda_{v}$ et $\mu_{v}$ des coefficients viscoélastiques.

\section{Application à une lame en flexion}
\subsection{Système d'équations en statique}
\medskip
Nous avons appliqué ce modèle au cas d'une lame de polyélectrolyte se déformant sous l'action d'un champ électrique 
permanent (cas statique). Les vitesses des différents constituants et les dérivées partielles par rapport au temps 
sont donc nulles. Nous avons étudié une lame de Nafion $Li^{+}$ de longueur $L=2~cm$, d'épaisseur $2e=200~\mu m$ et 
de largeur $2l=5~mm$ soumise à une différence de potentiel $\varphi_{0} = 1~V$. La charge massique $Z_{3}$ et la 
masse volumique $\rho _{3}^{0}$ de la phase solide sont les suivantes :
\begin{equation}
\begin{tabular}{ccc}
$Z_{3} \approx -9~10^{4}~C~kg^{-1} \cite{Collette} \qquad \qquad $ & $ \rho_{3}^{0} \approx 2078~kg~m^{-3}$  
$\cite{nemat2000}$%
\end{tabular}
\end{equation}

\begin{wrapfigure}{l}{0.6\textwidth}
\includegraphics [width=\linewidth]{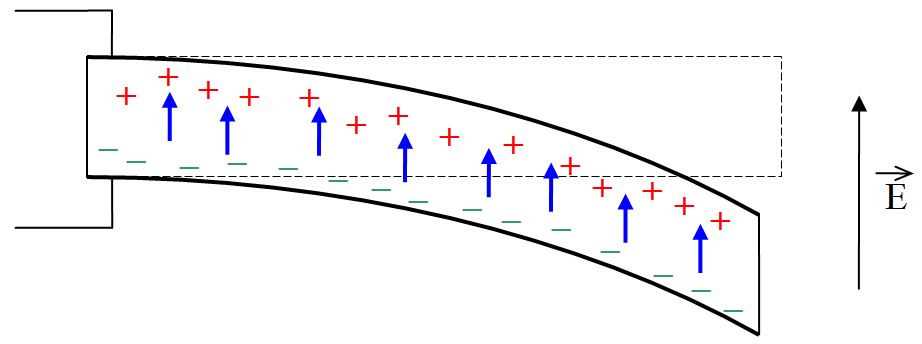}
\caption{Lame de PEA en flexion}
\label{fig:1}
\end{wrapfigure}

La fraction volumique de la solution $\phi_{4}$ est de l'ordre de $38\%$ \cite{nemat2000}. Nous admettrons 
dans ce qui suit qu'elle est uniforme dans tout le matériau. La température absolue est ${T=300~K}$, et la masse 
volumique de l'eau ${\rho_{2}^{0}=1000~kg~m^{-3}}$. La permittivité diélectrique du matériau a été mesurée par 
Deng et al \cite{Deng} pour un matériau très voisin du Nafion : ${\varepsilon \sim 10^{-6}~F~m^{-1}}$. 
Nous la considèrerons comme une constante.
Compte tenu des dimensions de la lame, le problème peut être considéré comme bidimensionnel. On choisit un 
repère $Oxyz$ tel que l'axe $Oz$ soit orthogonal à la lame  non déformée et parallèle au champ électrique 
imposé, l'axe $Ox$ suivant la longueur de la lame et l'axe $Oy$ suivant la largeur. On suppose en 
outre que la composante axiale $E_{x}$ du champ électrique induit à l'intérieur de la poutre est négligeable 
devant la composante normale $E_{z}$. Désignons par $\varphi$ le potentiel électrique. On admet qu'en première 
approximation, $C$, $E_{z}$, $p$, $\varphi$ et la charge électrique volumique $\rho Z$ ne dépendent que de la 
variable $z$. En supposant le terme de pression de l'équation (\ref{Nernst}) négligeable, le système d'équations 
s'écrit en projection :
\begin{equation}
\begin{tabular}{cc}
$E_{z}= - \frac {d \varphi} {dz} \qquad \qquad \qquad $ & 
$\varepsilon \frac {d E_{z}} {dz} = \rho Z $ \\
$\frac {dp} {dz} = \left(C F - \rho _{2}^{0} Z_{3}\right) E_{z} \qquad \qquad \qquad $ &
$\frac {dC} {dz} = \frac{FC}{RT} E_{z} $
\end{tabular} \label{Eqp}
\end{equation}%
avec :
\begin{equation}
\rho Z = \phi_{4} F \left(C - C_{moy} \right)
\end{equation}
où $C_{moy} = - \frac {\left(1-\phi_{4} \right) \rho _{3}^{0} Z_{3}} {\phi_{4} F} = 3082~mol~m^{-3}$ désigne 
la concentration moyenne en cations. Les conditions aux limites et la condition d'électroneutralité s'écrivent :
\begin{equation}
\begin{tabular}{ccc}
$\underset{z \rightarrow -e}{lim} \varphi =\varphi_{0}\qquad \qquad $ & $%
\underset{z \rightarrow e}{lim} \varphi = 0 \qquad \qquad $ & $
\int_{-e}^{e} \rho Z \, \mathrm{d}z =0 $
\end{tabular} \label{CL}
\end{equation}
 
Cette dernière condition équivaut à $E_{z}\left( e \right) = E_{z}\left( -e \right)$ d'après (\ref{Eqp}). On 
introduit les variables adimensionnées suivantes :
\begin{equation}
\begin{tabular}{ccc}
$\overline{E} = \frac {E_{z} e} {\varphi_{0}} \qquad \qquad $ & $ \overline{C} = \frac {C} {C_{moy}} \qquad 
\qquad $ & $ \overline{\varphi} = \frac {\varphi} {\varphi_{0}}$ \\
$\overline{\rho Z} = \frac {\rho Z} {\phi_{4} F C_{moy}} \qquad \qquad $ & $ 
\overline{p} = \frac {p} {F \varphi_{0} C_{moy}} \qquad \qquad $ & $ \overline{z} = \frac {z} {e}$ \\
\end{tabular}
\end{equation}
Les équations deviennent :
\begin{equation}
\overline{E}= - \frac {d \overline{\varphi}} {d \overline{z}} \label{M1a}
\end{equation}%
\begin{equation}
\frac {d \overline{E}} {d \overline{z}} = \frac {A_{1}} {A_{2}}\overline{\rho Z} \label{M2a}
\end{equation}%
\begin{equation}
\frac {d \overline{p}} {d\overline{z}} = \left(\overline{C} + A_{3} \right) \overline{E} \label{LDa}
\end{equation}%
\begin{equation}
\frac {d \overline{C}} {d \overline{z}} = A_{2} \overline{C} \overline{E}  \label{LNa}
\end{equation}%
\begin{equation}
\overline{\rho Z} = \overline{C} - 1 \label{6a}
\end{equation}
où les constantes sans dimension $A_{1}$, $A_{2}$ et $A_{3}$ valent :
\begin{equation}
\begin{tabular}{ccc}
$A_{1}=\frac {\phi_{4} e^{2} F^{2} C_{moy}} {\varepsilon R T} \sim 4,37~10^{7} \qquad $ &
$A_{2}=\frac { F \varphi_{0}} {R T} \sim 38,7 \qquad $ &
$A_{3}= -\frac {\rho_{2}^{0} Z_{3} C_{moy}} {F C_{moy}} \sim 0,303 $
\end{tabular}
\end{equation}
et avec les conditions aux limites :
\begin{equation}
\begin{tabular}{ccc}
$\underset{\overline{z} \rightarrow -1}{lim} \overline{\varphi} = 1\qquad \qquad $ & $%
\underset{\overline{z} \rightarrow 1}{lim} \overline{\varphi} = 0 \qquad \qquad $ & $
\overline{E}\left( 1 \right) = \overline{E}\left( -1 \right) $
\end{tabular} \label{CLa}
\end{equation}

\subsection{Résolution}
\medskip
En combinant les équations (\ref{M2a}), (\ref{LNa}) et (\ref{6a}), on obtient l'équation :
\begin{equation}
\frac {d} {d\overline{z}} \left( \frac {d \overline{C}} {\overline{C} d \overline{z}} \right) 
= A_{1} \left( \overline{C} - 1  \right) \label{E2bis}
\end{equation}
Par ailleurs :
\begin{equation}
\begin{tabular}{ccc}
$\overline{C} = A_{2} exp \left(-A_{2} \overline{\varphi} \right) \qquad \qquad $ & et  
$ \qquad \qquad \overline{p} = \frac {\overline{C}} {A_{2}} - A_{3} \overline{\varphi} $
\end{tabular} \label{E5E6}
\end{equation}
à une constante additive près.

La lame de polymère peut être assimilée à un matériau conducteur. On en déduit que le champ électrique est nul 
dans toute la lame excepté près des bords. Les valeurs des différents paramètres au centre de la lame et aux 
extrémités peuvent être déduites du système d'équation et conditions aux limites précédents  :
\begin{equation*}
\begin{tabular}{|c|c|c|c|}
\hline
& $-1$ & $0$ & $1$ \\ \hline
$\overline{C}$ & $A_{2}e^{-A_{2}}\simeq 0$ & $1$ & $A_{2}$ \\ \hline
$y$ & $\ln A_{2}-A_{2}$ & $0$ & $\ln A_{2}$ \\ \hline
$\overline{\varphi }$ & $1$ & $\frac{\ln A_{2}}{A_{2}}$ & $0$ \\ \hline
$\overline{E}$ & $\sqrt{\frac{2A_{1}}{A_{2}}\left[ 1-\frac{1}{A_{2}}\left( 1+\ln A_{2}\right)\right] }$ & $0$ & 
$\sqrt{\frac{2A_{1}}{A_{2}}\left[ 1-\frac{1}{A_{2}}\left( 1+\ln A_{2}\right)\right] }$ \\ \hline
$\overline{p}$ & $-A_{3}$ & $\frac{1}{A_{2}}\left( 1-A_{3}\ln
A_{2}\right) $ & $1$ \\ \hline
$\overline{\rho Z}$ & $-1$ & $0$ & $A_{2}-1$ \\ \hline
\end{tabular}%
\end{equation*}

L'équation (\ref{E2bis}) peut être résolue sous Matlab. On en déduit $\overline{E}$ par (\ref{LNa}), 
$\overline{\rho Z}$ par (\ref{6a}), et $\overline{\varphi}$ et $\overline{p}$ par (\ref{E5E6}).

Une évaluation du terme de pression de l'équation (\ref{Nernst}) montre que celui-ci n'excède pas $2\%$ du second 
terme de l'équation dans les conditions nominales choisies; l'erreur commise est proportionnelle au potentiel 
imposé $\varphi_{0}$ et reste inférieure à $10\%$ lorsque $\varphi_{0}=4~V$.

\subsection{Modèle de poutre en grands déplacements}
\medskip
Nous avons utilisé un modèle de poutre pour déterminer les efforts, contraintes et déplacements de la lame de PEA.
La poutre est encastrée à son extrémité $O$. L'autre extrémité $A$ est soit libre, soit soumise à un effort 
tranchant $\overrightarrow{F^{p}}$. Lorsqu'un champ électrique est appliqué, les cations et le solvant se 
déplacent vers l'électrode négative, entraînant une variation de volume et un fléchissement de la lame.
Le polymère est donc soumis à une force électrique appliquée aux cations et anions, que l'on peut modéliser par 
un effort réparti $\overrightarrow{p^{p}}$.

\begin{wrapfigure}{l}{0.5\textwidth}
\includegraphics [width=\linewidth]{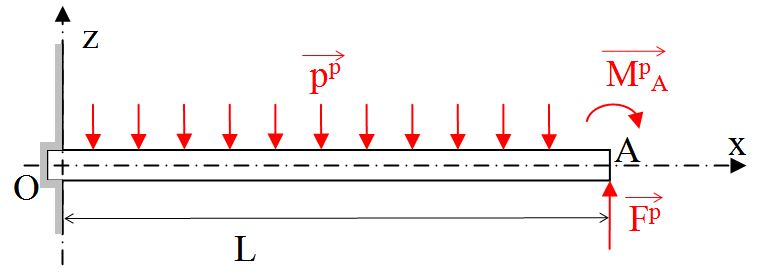}
\caption{Efforts exercés sur la poutre}
\label{fig:2}
\end{wrapfigure}

Compte tenu des hypothèses précédentes, cet effort réparti est 
indépendant de la coordonnée $x$ et est orthogonal à la lame. 
Par ailleurs, le gonflement de la lame du côté de l'électrode négative et sa contraction du côté opposé 
engendrent un moment fléchissant $\overrightarrow{M_{A}^{p}}$ à l'extrémité libre $A$ de la poutre.
L'effort réparti peut être calculé à l'aide de la relation suivante :
\begin{equation}
p^{p} = \int_{-l}^{l} \int_{-e}^{e} \rho Z E_{z}~dz~dy = 2l \int_{-e}^{e} \rho Z E_{z}~dz
= 2l \varepsilon \left[ \frac{E_{z}^{2}}{2} \right] _{-e}^{e} = 0
\end{equation}
compte tenu de (\ref{Eqp}) et de la condition d'électroneutralité. Le moment fléchissant est exercé suivant 
l'axe $Oy$ et résulte des efforts de pression $p = \frac {\sigma_{xx}} {3}$ : 
\begin{equation}
M_{A}^{p}= \int_{-l}^{l} \int_{-e}^{e}\sigma_{xx}~z~dz~dy = 6l \int_{-e}^{e} p~z~dz
\end{equation}

\begin{wrapfigure}{r}{0.45\textwidth}
\includegraphics [width=\linewidth]{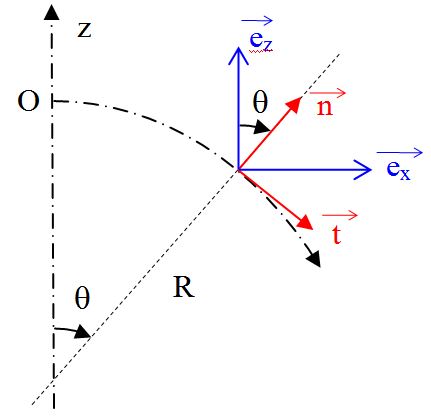}
\caption{Poutre en grands déplacements}
\label{fig:3}
\end{wrapfigure}

La flèche de la poutre pouvant atteindre des valeurs considérables, il convient de faire un calcul en grands 
déplacements. On admet que les sections droites de la lame restent planes et normales aux fibres après 
déformation (hypothèse de Bernoulli) et que les répartitions des contraintes et des 
déformations sont indépendantes des points d'application des forces extérieures (hypothèse de Barré Saint Venant).
Notons $s$ et $\overline{s}$ les abscisses curvilignes respectivement le long de la poutre 
au repos et de la poutre déformée, $\overrightarrow{t}$ et $\overrightarrow{n}$ les vecteurs tangent et normal 
à la poutre et $\theta$ l'angle de rotation d'une section droite. Aucun effort normal n'étant appliqué, 
on admettra que l'allongement $\Lambda = \frac{d \overline{s}}{ds}$ est égal à $1$. Le moment de flexion en 
une section quelconque vaut :
\begin{equation}
M^{p} = F^{p} \left( L-x \right) + M_{A}^{p}
\end{equation}

Le rayon de courbure $R$ est donné par :
\begin{equation}
\frac{1}{R} = \frac{d \theta}{d \overline{s}}
\end{equation}
Soit $\overrightarrow{u}$, le déplacement. Son gradient par rapport au repère $Oxyz$ lié à la poutre non déformée
vaut :
\begin{equation}
\overrightarrow{\overrightarrow{Grad}} \overrightarrow{u} = 
\begin{pmatrix}
   \left( 1 - \frac{\overline{n}}{R} \right) cos \theta - 1 & -sin \theta \\
   \left( 1 - \frac{\overline{n}}{R} \right) sin \theta & cos \theta - 1 
\end{pmatrix}
\end{equation}
où $\overline{n}$ désigne la coordonnée suivant la direction $\overrightarrow{n}$. Le tenseur des déformations 
est donné par :
\begin{equation}
\overrightarrow{\overrightarrow{\epsilon}} = \frac{1}{2} \left[ \left( \overrightarrow{\overrightarrow{Grad}} 
\overrightarrow{u} + \overrightarrow{\overrightarrow{1}} \right) ^{T} \left( 
\overrightarrow{\overrightarrow{Grad}} \overrightarrow{u} + \overrightarrow{\overrightarrow{1}} \right) 
- \overrightarrow{\overrightarrow{1}} \right]
\end{equation}
où $\overrightarrow{\overrightarrow{1}}$ désigne le tenseur identité. Il vient :
\begin{equation}
\epsilon_{xx} = - \frac{\overline{n}}{R} \left( 1 - \frac{\overline{n}}{2R} \right) \simeq - 
\frac{\overline{n}}{R}
\end{equation}
En effet, la poutre étant mince, $\left| \overline{n} \right| << R$.
Dans le cas d'une poutre en flexion pure, la déformation vaut $\epsilon_{xx}=\frac{M^{p}}{EI^{p}} \overline{n}$ 
où $E$ est le module d'Young et $I^{p}=\frac{4le^{3}}{3}$ le moment quadratique par rapport à l'axe $Oy$. L'effort 
tranchant a un effet négligeable sur la flèche. On en déduit :
\begin{equation}
\begin{tabular}{ccc}
$\frac{1}{R} = \frac{d \theta}{d \overline{s}} = \frac{F^{p}}{E I^{p}}  \left( L-\overline{s} 
\right) - \frac{M_{A}^{p}}{E I^{p}} \qquad \qquad $ & soit
$\qquad \qquad \theta = \frac{F^{p}}{2E I^{p}} \overline{s} \left( 2L-\overline{s} \right) - 
\frac{M_{A}^{p}}{E I^{p}} \overline{s} $
\end{tabular}
\end{equation}

en choisissant le point $O$ comme origine des abscisses curvilignes. La flèche $w$ est obtenue en intégrant la 
relation $\frac{dz}{d \overline{s}} = sin \theta$.

Dans le cas d'une poutre encastrée libre ($F^{p}=0$), le rayon de courbure est constant; la poutre prend la forme 
d'un arc de cercle et à l'extrémité :
\begin{equation}
\begin{tabular}{cc}
$\theta = - \frac{M_{A}^{p}}{E I^{p}} L \qquad \qquad $ &
$w = \frac{E I^{p}}{M_{A}^{p}} \left[ cos \left( \frac{M_{A}^{p}}{E I^{p}} L \right) -1 \right] $
\end{tabular}
\end{equation}

Lorsqu'une force de blocage est exercée à l'extrémité de la poutre, la flèche est nulle :
\begin{equation}
w = \int_{0}^{L} sin \left[\frac{F^{p}}{2E I^{p}} x \left( 2L-x \right) - \frac{M_{A}^{p}}{E I^{p}} 
x \right]~dx = 0
\end{equation}
Cette intégrale peut être calculée en utilisant les fonctions de Fresnel $S$ et $C$ :
\begin{equation}
\begin{tabular}{lll}
$S(x)$ & $=\int_{0}^{x}\sin t^{2}\;dt$ & $=\sum\limits_{n=0}^{+\infty} 
\left( -1\right) ^{n}\frac{x^{4n+3}}{\left( 2n+1\right) !\left( 4n+3\right) }$ \\ 
$C(x)$ & $=\int_{0}^{x}\cot s^{2}\;dt$ & $=\sum\limits_{n=0}^{+\infty}
\left( -1\right) ^{n}\frac{x^{4n+1}}{\left( 2n\right) !\left( 4n+1\right) }$%
\end{tabular}%
\end{equation}
Posons :
\begin{equation}
\begin{tabular}{lll}
$x^{\ast }=\sqrt{\frac{2EI^{p}}{F^{p}}}\qquad \qquad $ & $x_{0}=\frac{%
M_{A}^{p}}{F^{p}}-L\qquad \qquad $ & $x_{1}=\frac{M_{A}^{p}}{^{p}}$%
\end{tabular}%
\end{equation}
On obtient :
\begin{equation}
x^{\ast }\cos \left( \frac{x_{0}^{2}}{x^{\ast ^{2}}}\right) \left[
S\left( \frac{x_{1}} {x^{\ast }} \right) -S\left(  \frac{x_{0}} {x^{\ast }} \right) \right]
-x^{\ast }\sin \left( \frac{x_{0}^{2}}{x^{\ast ^{2}}}\right) \left[ C\left(
 \frac{x_{1}} {x^{\ast }} \right) -C\left(  \frac{x_{0}} {x^{\ast }} \right) \right] = 0
\end{equation}
Un calcul en petits déplacements montre que $F^{p} \simeq \frac{3 M_{A}^{p}}{2L}$. Compte tenu de la valeur de 
$M_{A}^{p}$ fournie par les simulations, $\frac{x_{0}}{x^{\ast }} < \frac{x_{1}}{x^{\ast }} << 1$. On peut 
donc remplacer $S$ et $C$ par les premiers termes de leurs développements limités, l'erreur commise 
étant inférieure à $1\%$. On obtient alors pour la force de blocage le même résultat qu'en petits déplacements :
\begin{equation}
F^{p} = \frac{3 M_{A}^{p}}{2L}
\end{equation}

On obtient en définitive les résultats suivants en variables adimensionnées :
\begin{equation}
M_{A}^{p}=A_{5}\int_{-1}^{1}\overline{p}\;\overline{z}\;d\overline{z} \label{MA}
\end{equation}
avec, pour la poutre encastrée-libre :
\begin{equation}
\begin{tabular}{cc}
$w = \frac{L^{2}}{2A_{7}\int_{-1}^{1}\overline{p}\;\overline{z}\;d \overline{z}}\left[ \cos \left( 
\frac{2A_{7}}{L}\int_{-1}^{1}\overline{p}\;\overline{z}\;d\overline{z}\right) -1\right] \qquad \qquad $ &
$\theta = -\frac{2A_{7}}{L}\int_{-1}^{1}\overline{p}\;\overline{z}\;d \overline{z} $
\end{tabular} \label{wtheta}
\end{equation}
et pour la force de blocage :
\begin{equation}
F^{p}=\frac{3A_{5}}{2L}\int_{-1}^{1} \overline{p}\;\overline{z}\;d\overline{z} \label{Fp}
\end{equation}%
où :
\begin{equation}
\begin{tabular}{cc}
$A_{5}=6le^{2}F\varphi _{0}C_{moy}\sim 0,045\;N\;m \qquad \qquad $ &  
$A_{7}=\frac{9}{4}\frac{L^{2}F\varphi _{0}C_{moy}}{eE}\sim 20,59\;m$%
\end{tabular} \label{A5A7}
\end{equation}

On peut maintenant vérifier que les variations de la fraction volumique $\phi_{4}$ sont négligeables. 
Considérons un petit élément de volume $dV$ situé à une distance $z$ de l'axe de la poutre. D'après l'hypothèse 
de Bernoulli, lorsque la poutre fléchit avec un rayon de courbure $R$, ce volume devient $\frac{\left| R 
\right|+z}{\left| R \right|} dV$. Le volume de la phase solide est invariable, seul le volume de la phase 
liquide change. La variation de fraction volumique de la phase liquide est de l'ordre de $\frac {\phi_{3} z} 
{\left| R \right|}$, soit une variation inférieure à $0,3\%$ sur l'épaisseur de la poutre.

\section{Résultats}

Les simulations ont été faites pour une lame de Nafion 117 $Li^{+}$ de dimensions nominales $L=2~cm$, 
$e=100~\mu m$ et $l=2,5~mm$ soumise à une différence de potentiel $\varphi_{0} = 1~V$. Les valeurs de permittivité
diélectrique relevées dans la littérature étant très dispersées, nous avons tout d'abord fait varier
ce paramètre entre $10^{-1}~F~m^{-1}$ et $10^{-11}~F~m^{-1}$. Les angles de rotation obtenus 
ne sont réalistes que pour des permittivités comprises entre $10^{-8}~F~m^{-1}$ et $10^{-5}~F~m^{-1}$. En 
comparant les valeurs obtenues pour la flèche et pour la force de blocage avec les valeurs expérimentales 
\cite{Nemat2002}, \cite{Newbury2002}, \cite{Newbury2003} et \cite{NewburyTh}, on obtient une valeur de $\varepsilon$ 
voisine de $10^{-6}~F~m^{-1}$, valeur proche de celle mesurée par Deng et al \cite{Deng}.

\begin{figure*} [h]
 \includegraphics[width=0.9\textwidth]{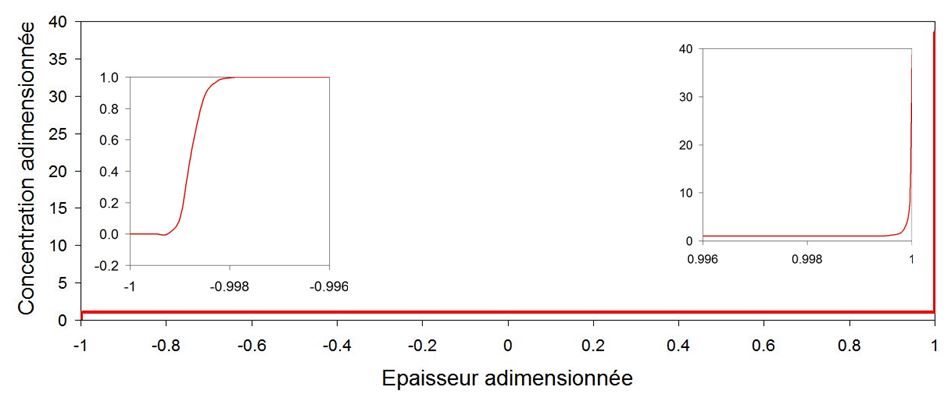}
\caption{Variation de la concentration en cations dans l'épaisseur de la lame}
\label{fig:4}
\end{figure*}

\subsection{Profils des différentes grandeurs}
\begin{figure*} [h]
 \includegraphics[width=0.9\textwidth]{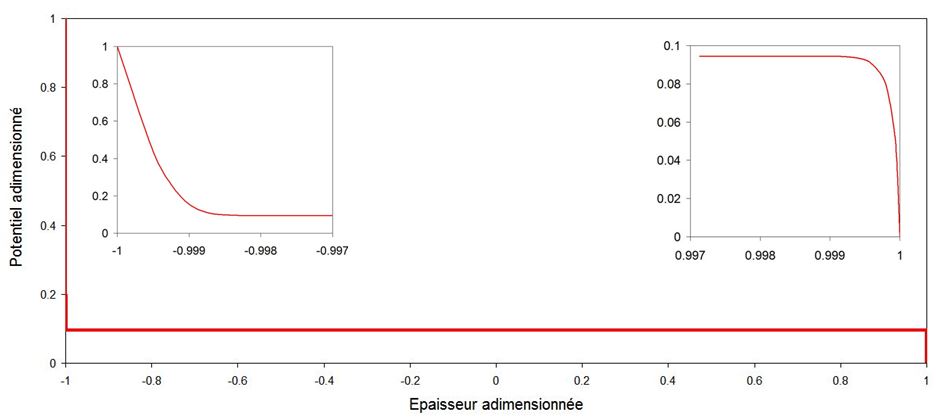}
\caption{Variation du potentiel électrique dans l'épaisseur de la lame}
\label{fig:5}
\end{figure*}

\begin{figure*} [h]
 \includegraphics[width=0.9\textwidth]{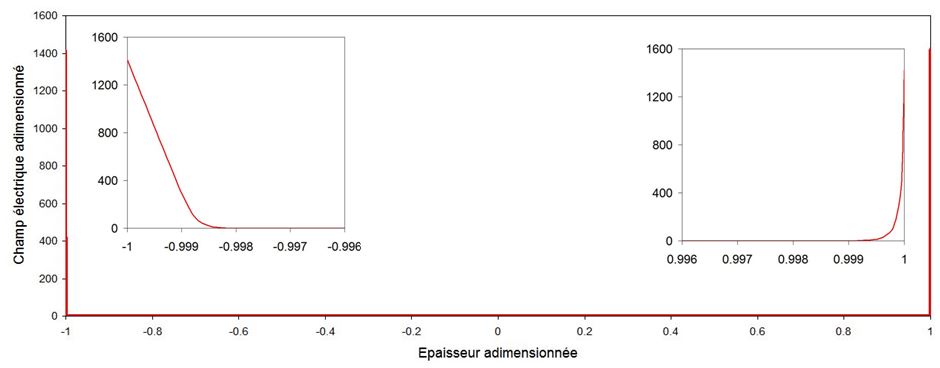}
\caption{Variation du champ électrique dans l'épaisseur de la lame}
\label{fig:6}
\end{figure*}

\begin{figure*} [!]
 \includegraphics[width=0.9\textwidth]{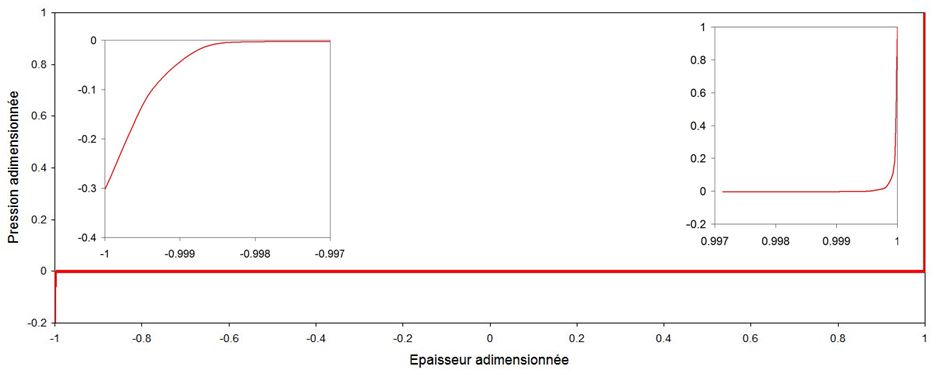}
\caption{Variation de la pression dans l'épaisseur de la lame}
\label{fig:7}
\end{figure*}

\medskip
Les variations de la concentration en cations, du potentiel électrique, du champ électrique et de la pression 
dans l'épaisseur de la lame sont représentées sur les figures \ref{fig:4}, \ref{fig:5}, \ref{fig:6} et \ref{fig:7}. 
Ces grandeurs sont constantes dans toute la partie centrale de la lame, mais varient très fortement au voisinage des 
électrodes, en particulier près de l'électrode négative sur laquelle s'accumulent les cations; a contrario, il existe 
à proximité de l'électrode positive une zone de l'ordre de $0,1~\mu m$ totalement dépourvue de cations.

\subsection{Influence de la tension entre les électrodes}
\begin{figure*} [h]
 \includegraphics[width=0.45\textwidth]{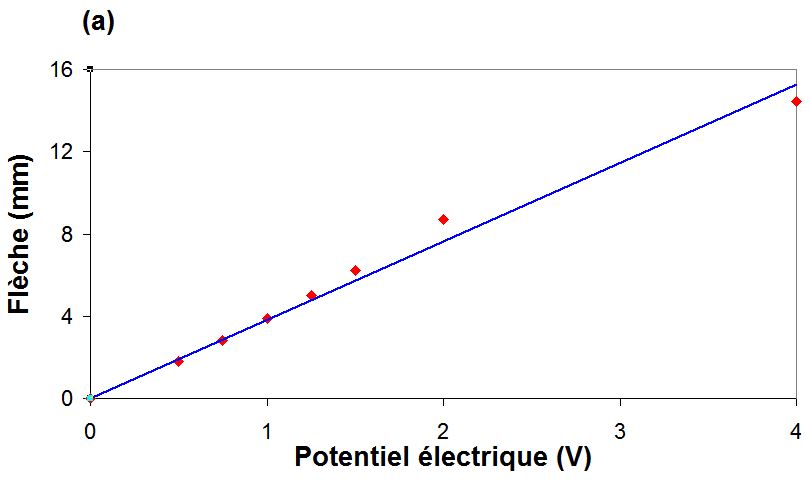}
 \qquad
 \includegraphics[width=0.45\textwidth]{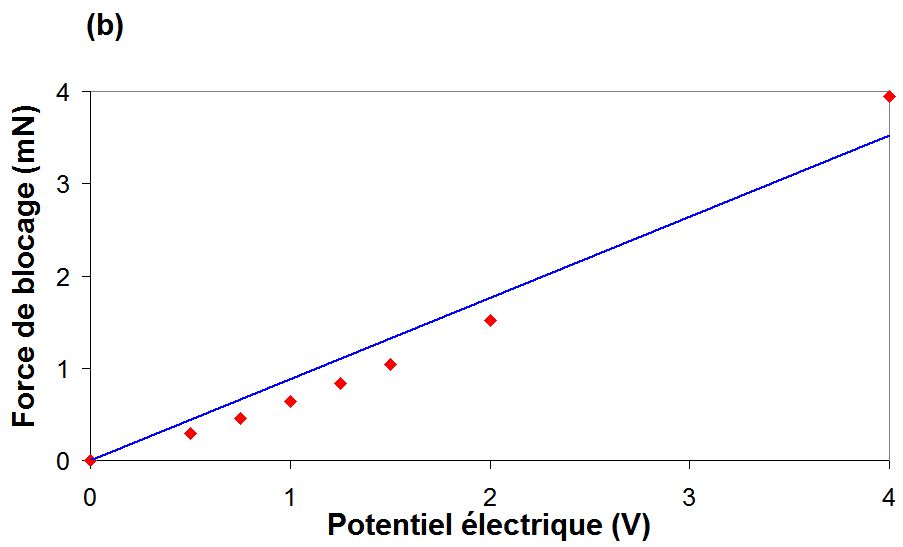}
\caption{Influence du potentiel électrique : (a) Sur la flèche (b) Sur la force de blocage }
\label{fig:8}
\end{figure*}

\medskip
Nous avons fait varier le potentiel imposé entre $0,5~V$ et $4~V$. On vérifie que la flèche varie linéairement 
avec la différence de potentiel $\varphi_{0}$, résultat en accord avec les expériences de Mojarrad et al 
\cite{Mojarrad} et Shahinpoor et al \cite{Shahinpoor1998}. On observe que la force de blocage suit la même 
tendance (figure \ref{fig:8}).

\subsection{Lois d'échelle}
\begin{figure*} [h]
 \includegraphics[width=0.3\textwidth]{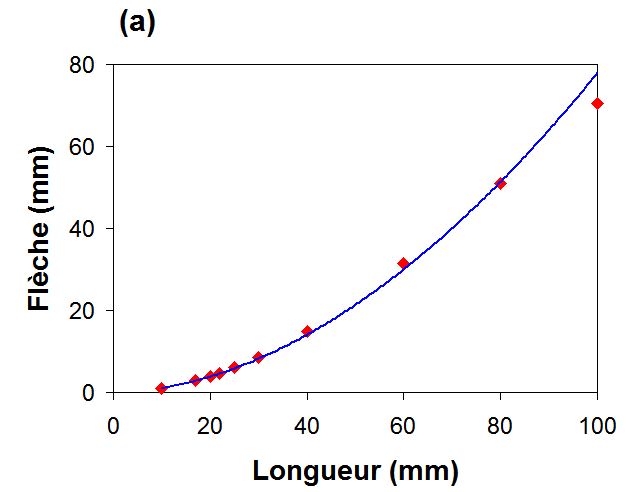}
 \qquad
 \includegraphics[width=0.3\textwidth]{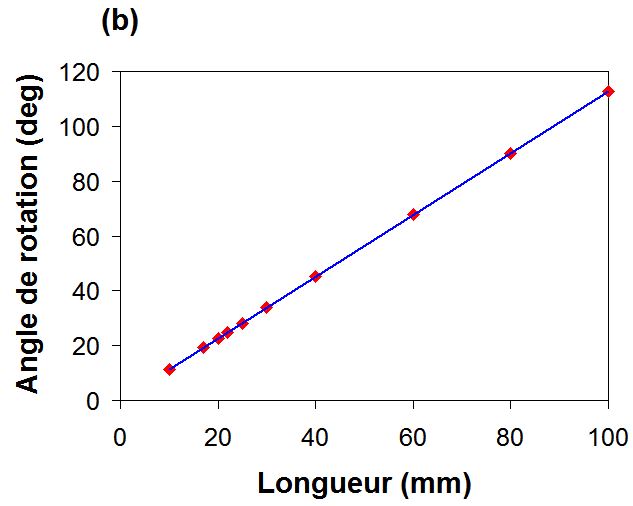}
  \qquad
 \includegraphics[width=0.3\textwidth]{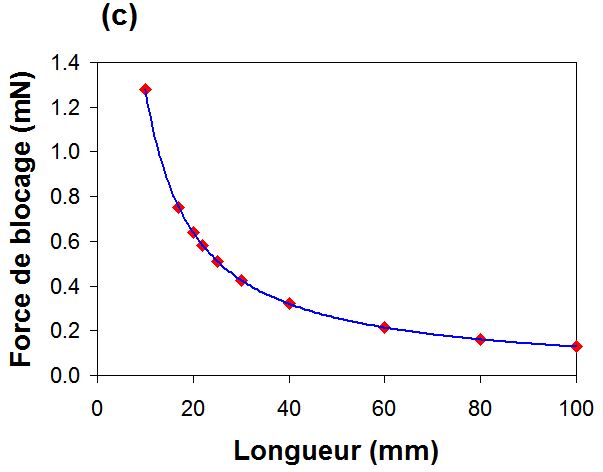}
\caption{Influence de la longueur : (a) Sur la flèche (b) Sur l'angle de rotation (c) Sur la force de blocage }
\label{fig:9}
\end{figure*}

\medskip
Nous avons également étudier l'influence de la géométrie de la poutre. D'après les équations obtenues, la flèche 
est indépendante de la largeur de la poutre. En effet le système d'équations utilisé pour 
calculer $\int_{-1}^{1}\overline{p}\;\overline{z}\;d \overline{z}$ ne dépend que du potentiel imposé $\varphi_{0}$,
de l'épaisseur $e$ et du matériau choisi. La constante $A_{7}$ étant indépendante de $l$, il en est de même 
de la flèche $w$. Quant à la constante $A_{5}$, elle varie linéairement avec la largeur, de même que la force de 
blocage. Ce résultat est en accord avec \cite{Newbury2003}. \newline
Nous avons fait varier la longueur de la lame entre $L=1~cm$ et $L=10~cm$ (figure \ref{fig:9}). D'après les relations 
\ref{wtheta}, l'angle de rotation est donc une fonction linéaire de $L$ tandis que la flèche est 
approximativement proportionnelle au carré de la longueur, résultat en bon accord avec les mesures de Shahinpoor 
\cite{Shahinpoor1999}. Quant à la force de blocage, elle est inversement proportionnelle à la longueur, ce qui est en 
accord avec les résultats expérimentaux de Newbury et al \cite{Newbury2003}. \newline

\begin{figure*}
 \includegraphics[width=0.45\textwidth]{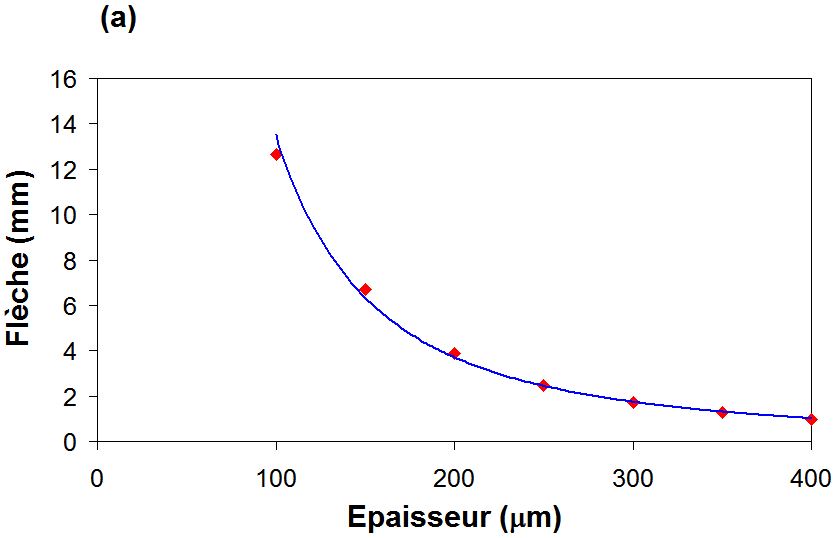}
 \qquad
 \includegraphics[width=0.45\textwidth]{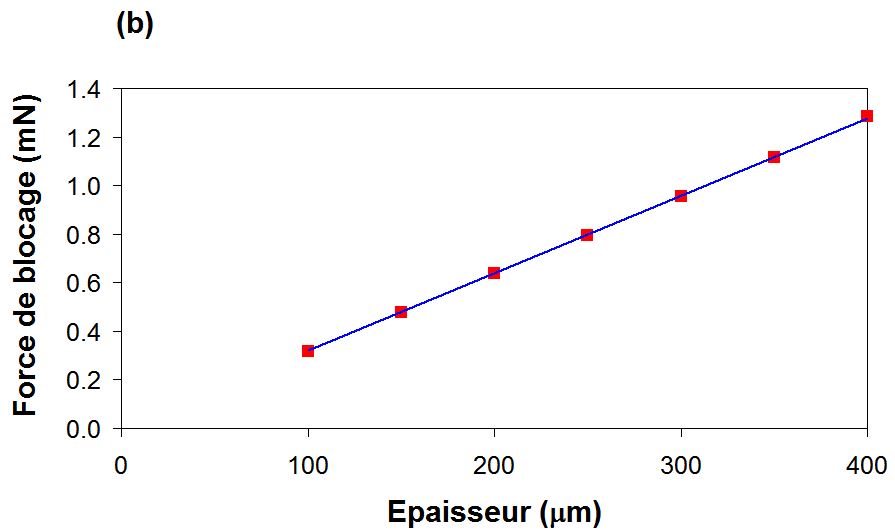}
\caption{Influence de l'épaisseur : (a) Sur la flèche (b) Sur la force de blocage }
\label{fig:10}
\end{figure*}

Nous avons enfin étudié l'influence de l'épaisseur  de la lame pour des valeurs comprises entre $50~\mu m$ et 
$400~\mu m$ (figure \ref{fig:10}). On observe que la force de blocage et le moment fléchissant $M_{A}^{p}$ varient linéairement avec $e$, et que la flèche et l'angle de rotation varient en $e^{-2}$.

\section{Conclusion}
\medskip
Nous avons étudié la flexion d'une lame de polyélectrolyte à l'aide d'un modèle issu de la thermodynamique des 
processus irréversibles que nous avions précédemment développé. Une différence de potentiel continue est appliquée
entre les deux faces de la lame (cas statique). L'une des extrémités est encastrée; l'autre extrémité est soit libre, 
soit soumise à une force ramenant la flèche à une valeur nulle (force de blocage). Nous avons calculé les efforts 
appliqués et la flèche à l'aide d'un modèle de poutre en grands déplacements. Le matériau choisi pour effectuer les 
simulations est le Nafion. \newline
Nous avons obtenu les profils de concentration en cations, de charge électrique, de potentiel et de champ électrique
dans l'épaisseur de la lame. On observe que ces grandeurs sont quasiment constantes dans toute la partie centrale de 
la lame et qu'elles varient très fortement au voisinage des électrodes, ce qui est caractéristique du comportement 
d'un matériau conducteur. Nous obtenons pour la flèche et de la force de blocage des valeurs en bon accord avec les 
données expérimentales publiées dans la littérature : ces deux grandeurs varient linéairement avec la différence de 
potentiel imposée; la flèche est indépendante de la largeur de la lame, proportionnelle à sa longueur au carré et 
varie comme l'inverse du carré de l'épaisseur. La force de blocage est proportionnelle à la largeur, varie comme 
l'inverse de la longueur et linéairement avec l'épaisseur. \newline
Une piste d'amélioration de ces résultats est de prendre en compte les variations de la permittivité avec la 
concentration en cations. Nous envisageons également d'étudier l'effet inverse (mode capteur) et d'autres 
configurations, notamment le cas d'une lame encastrée à ses deux extrémités.

\section{Notations}
\medskip
Les indices $k=1,2,3,4$ désignent respectivement les cations, le solvant, le solide et la solution. 
Les quantités non indicées sont relatives au matériau complet. L'exposant $^{s}$ désigne la partie symétrique 
sans trace d'un tenseur du second ordre.\newline
\newline
\noindent$C$ ($C_{moy}$) : concentration molaire en cations (relative à la phase
liquide);\newline
\noindent$D$ : coefficient de diffusion de masse des cations dans la phase
liquide;\newline
\noindent$\overrightarrow{D}$ : induction électrique;\newline
\noindent$e$ : demi épaisseur de la lame;\newline
\noindent$E$ ($U$) : énergie volumique totale (interne);\newline
\noindent$\overrightarrow{E}$ : champ électrique;\newline
\noindent$F$ : constante de Faraday;\newline
\noindent$\overrightarrow{F^{p}}$ : force de blocage;\newline
\noindent$G$, $\lambda $, $E$ : coefficients élastiques;\newline
\noindent$\overrightarrow{I}$ : densité volumique de courant;\newline
\noindent$I^{p}$ : moment quadratique de la poutre;\newline
\noindent$K$ : perméabilité intrinsèque de la phase solide;\newline
\noindent$l$ : demi largeur de la lame;\newline
\noindent$L$ : longueur de la lame;\newline
\noindent$M_{k}$ : masse molaire du constituant $k$;\newline
\noindent$\overrightarrow{M^{p}}$ ($\overrightarrow{M_{A}^{p}}$) : moment fléchissant; \newline
\noindent$p$ : pression;\newline
\noindent$\overrightarrow{Q}$ : flux de chaleur;\newline
\noindent$R$ : constante universelle des gaz parfaits;\newline
\noindent$S$ : entropie volumique;\newline
\noindent$T$ : température absolue;\newline
\noindent$v_{k}$ : volume molaire partiel du constituant $k$ (relatif à la phase liquide);\newline
\noindent$\overrightarrow{V}$ ($\overrightarrow{V_{k}}$) : vitesse;\newline
\noindent$w$ : flèche de la poutre;\newline
\noindent$Z$ ($Z_{k}$) : charge électrique massique;\newline
\noindent$\varepsilon $ : permittivité diélectrique;\newline
\noindent$\utilde{\epsilon }$ : tenseur des déformations;\newline
\noindent$\eta$ : viscosité dynamique de l'eau;\newline
\noindent$\theta$ : angle de rotation des sections droites de la poutre;\newline
\noindent$\lambda _{v}$\textit{, }$\mu _{v}$\textit{\ }: coefficients viscoelastiques;\newline
\noindent$\mu _{k}$ : potentiel chimique massique;\newline
\noindent$\rho $ ($\rho _{k}$, $\rho _{k}^{0}$) : masse volumique;\newline
\noindent$\utilde{\sigma }$ ($\utilde{\sigma ^{es}}$, $\utilde{\sigma_{k}}$) : tenseur des contraintes 
totales (d'équilibre);\newline
\noindent$\varphi$ ($\varphi_{0}$): potentiel électrique;\newline
\noindent$\phi _{k}$ : fraction volumique de la phase $k$;\newline


\begin{thebibliography}{9}
\bigskip
%
\bibitem{nemat2000} S. Nemat-Nasser, J. Li, Electromechanical response of
ionic polymers metal composites, Journal of Applied Physics, 87 (2000) 3321--3331.
%
\bibitem{Ishii06} M. Ishii, T. Hibiki, Thermo-fluid dynamics of two-phase
flow, Springer, New-York, 2006.
%
\bibitem{Tixier1} M. Tixier, J. Pouget, Conservation laws of an
electro-active polymer, Continuum Mechanics and Thermodynamics 26, 4 (2014) 465--481.
%
\bibitem{Tixier2} M. Tixier, J. Pouget, Constitutive equations for an electroactive polymer, Continuum 
Mechanics and Thermodynamics, 28, 4 (2016) 1071--1091.
%
\bibitem{Collette} F. Collette, Vieillissement hygrothermique du Nafion, Thèse, Université de Grenoble  I, 2008.
%
\bibitem{Deng} Z.D. Deng, K.A. Mauritz, Dielectric relaxation studies of water-containing short side chain 
perfluorosulfonic acid membranes, Macromolecules, 25 (1992) 2739--2745.
%
\bibitem{Nemat2002} S. Nemat-Nasser, Micromechanics of actuation of ionic polymer-metal composites, Journal 
of Applied Physics, 92, 5 (2002) 2899--2915.
%
\bibitem{Newbury2002} K.M. Newbury, D.J. Leo, Linear Electromechanical modeling and characterization of ionic 
polymer benders, Journal of Intelligent Material Systems and Structures, 13 (2002) 51--60.

\bibitem{Newbury2003} K.M. Newbury, D.J. Leo, Electromechanical Model of ionic polymer transducers - Part II : 
experimental validation, Journal of Intelligent Material Systems and Structures, 14 (2003) 343--357.

\bibitem{NewburyTh} K.M. Newbury, Characterization, modeling and control of ionic-polymer transducers, 
Thesis, Faculty of the Virginia Polytechnic Institute and State University, Blacksburg, Virginia, 2002.
%
\bibitem{Mojarrad} M. Mojarrad, M. Shahinpoor, Ion-exchange-metal composite sensor films, Proceedings of the 
SPIE, 3042 (1997) 52--60.
%
\bibitem{Shahinpoor1998} M. Shahinpoor, Y. Bar-Cohen, J.O. Simpson, J. Smith, Ionic polymer-metal composites
(IPMCs) as biomimetic sensors, actuators and artificial muscles; a review, Smart Materials and Structures 7, 
(1998) 15--30.
%
\bibitem{Shahinpoor1999} M. Shahinpoor, Electro-mechanics of iono-elastic beams as electrically-controllable 
artificial muscles, Proceedings of SPIE, 3669 (1999) 109--121.
%
\end{thebibliography}
\end{document}